\documentclass[manuscript]{aastex}

\shorttitle{Chandra Observations of SDSS~J1029+2623}
\shortauthors{Ota et al.}

\begin{document}

\title{The Chandra view of the Largest Quasar Lens SDSS~J1029+2623}

\author{
Naomi Ota,\altaffilmark{1}  
Masamune Oguri,\altaffilmark{2} 
Xinyu Dai,\altaffilmark{3} 
Christopher S. Kochanek,\altaffilmark{4} 
Gordon T. Richards,\altaffilmark{5}
Eran O. Ofek,\altaffilmark{6}
Roger D. Blandford,\altaffilmark{7} 
Tim Schrabback,\altaffilmark{8} and 
Naohisa Inada\altaffilmark{9} 
}

\altaffiltext{1}{Department of Physics, Nara Women's University, Kitauoyanishimachi, Nara, Nara 630-8506, Japan.}
\altaffiltext{2}{Kavli Institute for the Physics and Mathematics of the Universe, University of Tokyo, 5-1-5 Kashiwa-no-ha, Kashiwa, Chiba 277-8568, Japan.}
\altaffiltext{3}{Homer L. Dodge Department of Physics and Astronomy, University of Oklahoma, Norman, OK 73019, USA.}
\altaffiltext{4}{Department of Astronomy, The Ohio State University, Columbus, OH 43210, USA.}                 
\altaffiltext{5}{Department of Physics, Drexel University, 3141 Chestnut Street,  Philadelphia, PA 19104, USA.}
\altaffiltext{6}{Benoziyo Center for Astrophysics, Weizmann Institute of Science, 76100 Rehovot, Israel.}
\altaffiltext{7}{Kavli Institute for Particle Astrophysics and Cosmology, 2575 Sand Hill Rd., Menlo Park, CA 94309, USA.}
\altaffiltext{8}{Kavli Institute for Particle Astrophysics and Cosmology, 
Stanford University, 382 Via Pueblo Mall, Stanford, CA 94305-4060, USA}
\altaffiltext{9}{Department of Physics, Nara National College of Technology, Yamatokohriyama, Nara 639-1080, Japan.}    

\begin{abstract}
  We present results from {\it Chandra} observations of the cluster
  lens SDSS~J1029+2623 at $z_l=0.58$, which is a gravitationally lensed 
  quasar with the largest known image separation.  We clearly detect X-ray
  emission both from the lensing cluster and the three lensed quasar
  images. The cluster has an X-ray temperature of
  $kT=8.1^{+2.0}_{-1.2}$~keV and bolometric luminosity of $L_{\rm X} =
  9.6\times10^{44}~{\rm erg\,s^{-1}}$.  Its surface brightness is
  centered near one of the brightest cluster galaxies, and
  it is elongated East-West. We identify a subpeak North-West of the
  main peak, which is suggestive of an ongoing merger. Even so, the
  X-ray mass inferred from the hydrostatic equilibrium assumption
  appears to be consistent with the lensing mass from the Einstein
  radius of the system. We find significant absorption in the soft
  X-ray spectrum of the faintest quasar image, which can be
  caused by an intervening material at either the lens or source
  redshift. The X-ray flux ratios between the quasar images (after
  correcting for absorption) are in reasonable agreement with those at
  optical and radio wavelengths, and all the flux ratios are
  inconsistent with those predicted by simple mass models. This
  implies that microlensing effect is not significant for this system
  and dark matter substructure is mainly responsible for the anomalous
  flux ratios. 
\end{abstract}



\section{Introduction}

The gravitational lens \object{SDSS~J1029+2623} consists of three
images of a $z_s = 2.197$ quasar produced by a foreground cluster at
$z_l \approx 0.58$ \citep{inada06,oguri08}. With a maximum image
separation of 22\farcs5, this is the largest lens among $\sim 130$
gravitationally lensed quasars known to date. The lensing
interpretation is secure from multiple observational facts, including
the remarkably similar optical spectra with mini broad absorption line
(BAL) features in their emission lines, similar radio loudnesses
\citep{kratzer11}, and the identification of the lensing cluster in
the optical and with weak lensing \citep{oguri12}. Furthermore, this
lens system represents a rare example of a naked cusp lens. In this
configuration, only three images with similar brightnesses are
produced on the same side of the lens potential, which is very rare
among galaxy-scale quasar lenses \citep[e.g.,][]{lewis02} 
but has been predicted to be fairly
common among cluster-scale lenses \citep{oguri04b,li07,minor08}.

SDSS~J1029+2623 is one of two examples of large-separation quasar
lenses produced by massive clusters of galaxies. The other
large-separation lens is \object{SDSS~J1004+4112}
\citep{inada03,oguri04a,sharon05,inada05}, which consists of five
lensed images of a quasar at $z=1.734$ and a lensing cluster at
$z=0.68$. In fact, these clusters can be regarded as unique examples
of {\it strong-lens-selected} clusters in the sense that they have
been identified by searching for strong lenses in a large sample of
quasars. In particular, the uniqueness of the selection of these
lenses begs the questions of (1) do the strong-lens-selected clusters
follow empirical scaling relations derived from optical/X-ray selected
clusters? and (2) what is the dynamical state of these clusters?

{\it Chandra} and {\it XMM-Newton} observations of the other
large-separation quasar lens, SDSS~J1004+4112, have shown that the
lensing cluster is relaxed, with the X-ray temperature and luminosity
being consistent with the predictions of empirical scaling relations
within the observed scatter \citep{ota06b,lamer06}. This
interpretation is also supported by the recent comparison of the X-ray
measurements with detailed strong lens models \citep{oguri10}. There
is excellent agreement on the centroids and position angles between
the inferred dark matter distribution, the X-ray surface brightness,
and the brightest cluster galaxy.  On the other hand, the available
observational data suggest that the lensing cluster of SDSS~J1029+2623
may be complicated. For instance, the existence of two bright
elliptical galaxies at the center with a large velocity difference
($\sim 2800~{\rm km\,s^{-1}}$) between them indicates a possible
merger, although it can also be a chance projection of different halos
along the line-of-sight.  It may be that there is a possible
elongation of the mass distribution along line-of-sight, which can be
tested by comparing X-ray and lensing-derived masses.

In addition to the properties of the lensing cluster, the
high-resolution {\it Chandra} observations provide an additional probe
of the anomalous flux ratios between the quasar
images. \citet{oguri08} pointed out that observed optical flux ratios
of the three quasar images (named image A, B, and C) are markedly
different from what simple mass models predict. \citet{kratzer11}
found that the anomalous flux ratio persists in the radio image for
which the effects of differential dust extinction and microlensing by
stars are less important. In particular, there is a large flux
difference between the close quasar image pair near a fold caustic
(images B and C), which implies that a large part of the flux anomaly
is caused by (sub-)structure in the mass distribution of the lensing
cluster \citep{kratzer11}. However, there is a clear wavelength
dependence to the optical flux ratios \citep{oguri08}, which indicates
that chromatic microlensing or dust extinction must also affect the
third quasar image especially in optical.

In this paper, we present results from {\it Chandra} observations of
this unique cluster-scale quasar lens. We successfully detected X-ray
emission from both the lensing cluster and the three quasar images,
which are discussed in \S\ref{sec:cluster} and \S\ref{sec:quasar},
respectively. We compare our results with observations at other
wavelengths, as well as strong lensing information from the quasar
images. Throughout the paper we adopt a cosmological model with matter
density $\Omega_M=0.27$, cosmological constant $\Omega_\Lambda=0.73$,
and Hubble constant $H_0=70~{\rm km\,s^{-1}Mpc^{-1}}$. At the cluster
redshift of $z=0.584$, $1\arcsec$ corresponds to 6.69~kpc. Unless
otherwise specified, quoted errors indicate the 90\% confidence range.

\section{Observations}

\object{SDSS~J1029+2623} was observed for 60~ks with the {\it Chandra}
Advanced CCD Imaging Spectrometer \citep[ACIS;][]{gpg03} on 2010 March
11 (Observation ID: 11755). The data were obtained with the ACIS S3
CCD operating in VFAINT mode. ACIS-S was chosen because it has higher
sensitivity than ACIS-I and that the cluster emission is expected to
be {\it soft} due to the high redshift of the system.  This CCD has a
$1024\times1024$ pixel format with an image scale of
$0\farcs492$~pixel$^{-1}$.  The target was offset from the nominal aim
point with a Y-offset of $-1\arcmin$, which has little effect on the
spatial resolution.  The CCD temperature during the observations was
$-120{\rm ^{\circ}C}$.  The data were processed using the standard
software packages {\tt CIAO 4.3} and {\tt CALDB 4.4.1}. The background
was stable during the observation and the net exposure time after
applying standard light-curve screening cuts was 55815~s.

The ACIS image in the 0.5--7~keV band is shown in
Figure~\ref{fig:image}.  The extended cluster emission and the three
images of SDSS~J1029+2623 are all detected. The astrometry offset is
found to be negligibly small ($0\farcs2$) from the comparison of
quasar A position with the optical image. The cluster emission
exhibits a somewhat elongated feature, which we will study in detail
in \S\ref{subsec:cluster_image}. In the ACIS-S3 field, 42 point-like
sources, including the three quasar images, were detected with the
{\tt wavdetect} algorithm using a significance threshold
parameter\footnote{The sigthresh parameter is the significance
  threshold for source detection and is proportional to the inverse of
  the number of pixels in the image. A sigthresh parameter of
  $10^{-6}$ means that $\sim1$ false source per field is allowed.} of
$10^{-6}$.

\section{Lensing cluster}\label{sec:cluster}

\subsection{Spectral analysis}\label{subsec:cluster_spec}

We derive the spectrum of the cluster component by extracting the data
from a circular region with a radius of $1\arcmin$ centered on the
X-ray centroid, which is chosen so as to enclose about 90\% of source
photons and to avoid low signal-to-noise regions in the outskirts.  We
exclude $5''$ radius regions around each quasar image.  The background
was estimated in a surrounding annulus ($2\farcm2<r<2\farcm5$), rather
than using blank-sky observations because the background intensity
generally depends on the direction of the sky and time. The 0.5--7~keV
source count rate within $r<1\arcmin$ is $0.068\pm0.001$~${\rm
  counts\,s^{-1}}$, after the background subtraction.  In total, there
are $\sim$4800 net photons, whereas the background contribution to the
total spectrum is about 20\%. We fit the cluster spectrum in the
0.5--7~keV band with the APEC thin-thermal plasma model
\citep{smith01} utilizing {\tt XSPEC} version~12.6.0q.  The spectral
data was rebinned so that each bin contains more than 25 counts and
the chi-square statistics was utilized in the fitting.  The Galactic
hydrogen column density was fixed to $N_{\rm H}=1.67\times 10^{20}{\rm
  cm^{-2}}$ based on the LAB survey \citep{kalberla05}.  Given the
good agreement between the X-ray emission centroid and the position of
G2 galaxy (see below), we set the cluster redshift to that of G2
\citep[$z=0.584$;][]{oguri08}. We note that, when the hydrogen column
density or the cluster redshift is not fixed but is fitted to the
spectrum, the resulting values are consistent with those adopted here.

\begin{figure}
\plotone{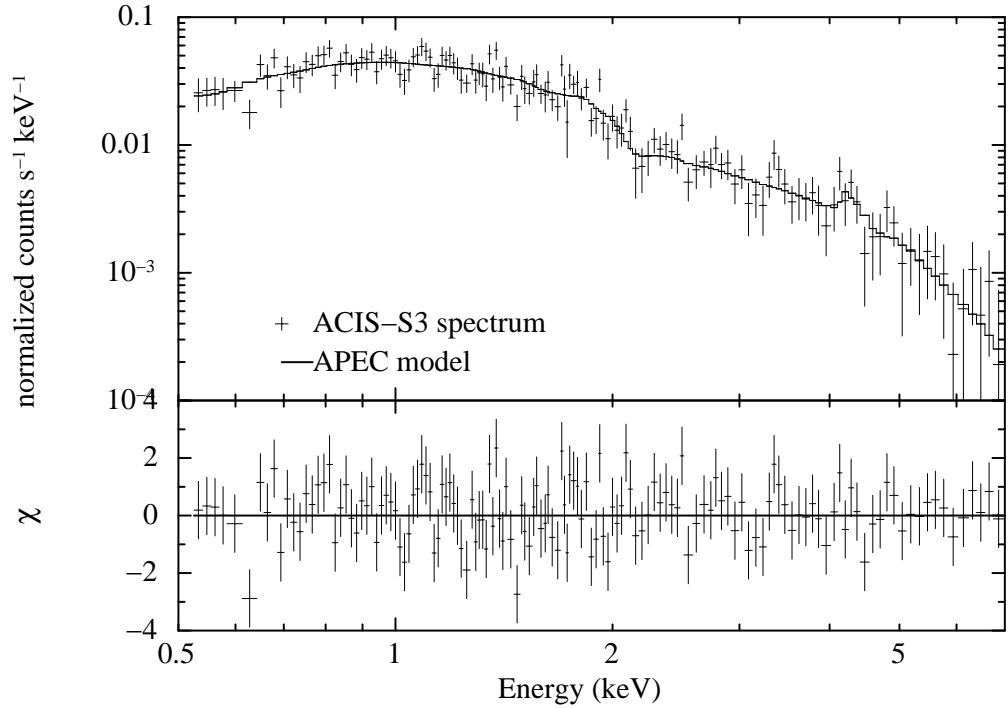}
\caption{{\it Chandra} ACIS-S3 spectrum of SDSS~J1029+2623
  ($r<1\arcmin$) fitted with the APEC model. In the top panel, the
  crosses denote the spectrum in the observed frame and the histogram function shows the
  best-fit model convolved with the telescope and 
  detector response functions. The bottom panel shows the residuals of
  the fit in units of $\sigma$. \label{fig:cluster_spec}} 
\end{figure}

The cluster spectrum is shown in Figure~\ref{fig:cluster_spec}. From
the spectral fits the X-ray temperature is constrained to be
$kT=8.1^{+2.0}_{-1.2}$~keV and the metal abundance to be
$Z=0.21Z_\odot$ (the 90\% upper limit is $0.44Z_\odot$).  This high
temperature indicates that the lensing cluster is indeed massive
enough to be capable of producing the large-separation lensed quasar
images (see also \S\ref{subsec:cluster_mass}). The reduced chi-square
of the best-fit model is $\chi^2/{\rm dof}=138/139$, and the
parameters of the fit are summarized in Table~\ref{tab:cluster_spec}.

The absorption-corrected 0.5--7~keV flux is $4.6\times10^{-13}{\rm
  erg\,s^{-1}}$ ($r<1\arcmin$).  The bolometric X-ray luminosity
within $r_{500}$ is estimated as $L_{\rm X}= 9.6\times10^{44}~{\rm
  erg\,s^{-1}}$, where $r_{500}$ is defined as the radius within which
the average matter density is equal to $\Delta_c=500$ times the
critical density of the Universe at the cluster redshift.  The
$\beta$-model analysis in \S\ref{subsec:cluster_image} yields
$r_{500}=1.09$~Mpc.  This luminosity is lower than the mean value
expected from the luminosity-temperature relation of distant clusters,
$L_{\rm X} = 4.0\times10^{45}~{\rm erg\,s^{-1}}$ \citep{ota06a}, but
it is within the observed scatter of the $L_{\rm X}-T$ relation.

To investigate the radial temperature profile, we also analyze spectra
for the inner $r<0\farcm25(\sim 100~{\rm kpc})$ region and the outer
$0\farcm25<r<1\arcmin$ region of the lensing cluster. The fitted
temperatures are $kT=9.8^{+3.7}_{-2.1}$~keV and
$6.8^{+1.7}_{-1.2}$~keV, respectively. While the temperature profile
is consistent with being constant with radius, the possible decrease
of the temperature toward large radii implies that the lensing cluster
lacks a cool core that is commonly seen in relaxed clusters
\citep{arnaud10}, although there is a possibility that a subpeak
discussed in \S\ref{subsec:cluster_image} is in fact a cool core of
this cluster.

\subsection{Image analysis}\label{subsec:cluster_image}

The X-ray centroid position (10:29:12.47, +26:23:33.2), measured in
\S\ref{subsec:cluster_image}, agrees well with the position of the
central galaxy G2 at (10:29:12.48, +26:23:32.0; \citealt{inada06}).
We first fit a single one-dimensional radial surface-brightness
profile to the extended X-ray emission. We centered the model at the
X-ray centroid and azimuthally averaged the 0.5--5~keV image. We
exclude $>5$~keV because the significance of X-ray emission above 5
keV is low (see Figure~\ref{fig:cluster_spec}). We corrected for
vignetting and the detector response using the exposure map calculated
for the spectral energy distribution of the cluster. The image was
rebinned by a factor of two so that the pixel scale is $0\farcs98$
(6.58~kpc at the cluster redshift).  The quasar images were masked out
using a $5\arcsec$ radius when calculating the profile.

\begin{figure}
\epsscale{1.1}
\plottwo{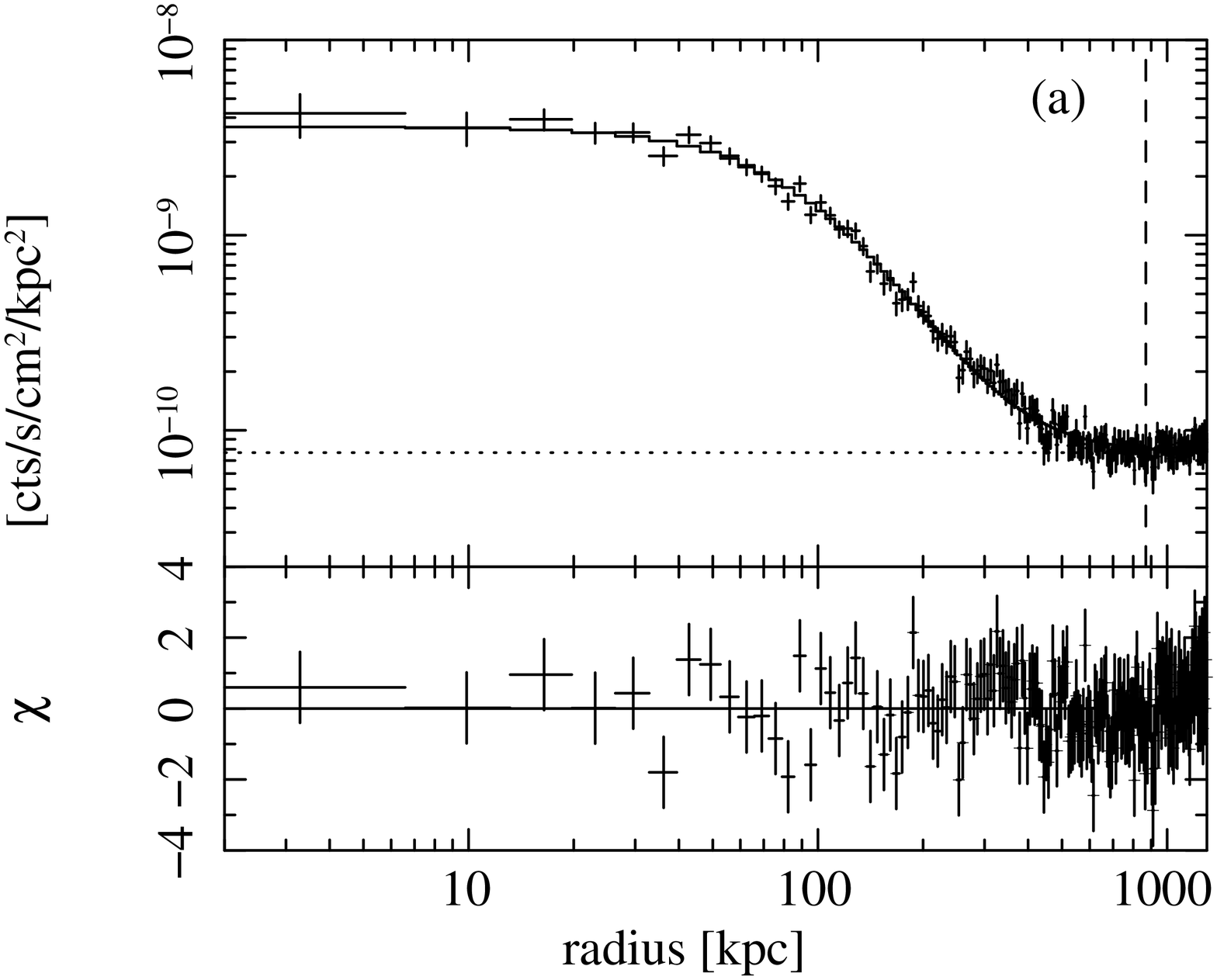}{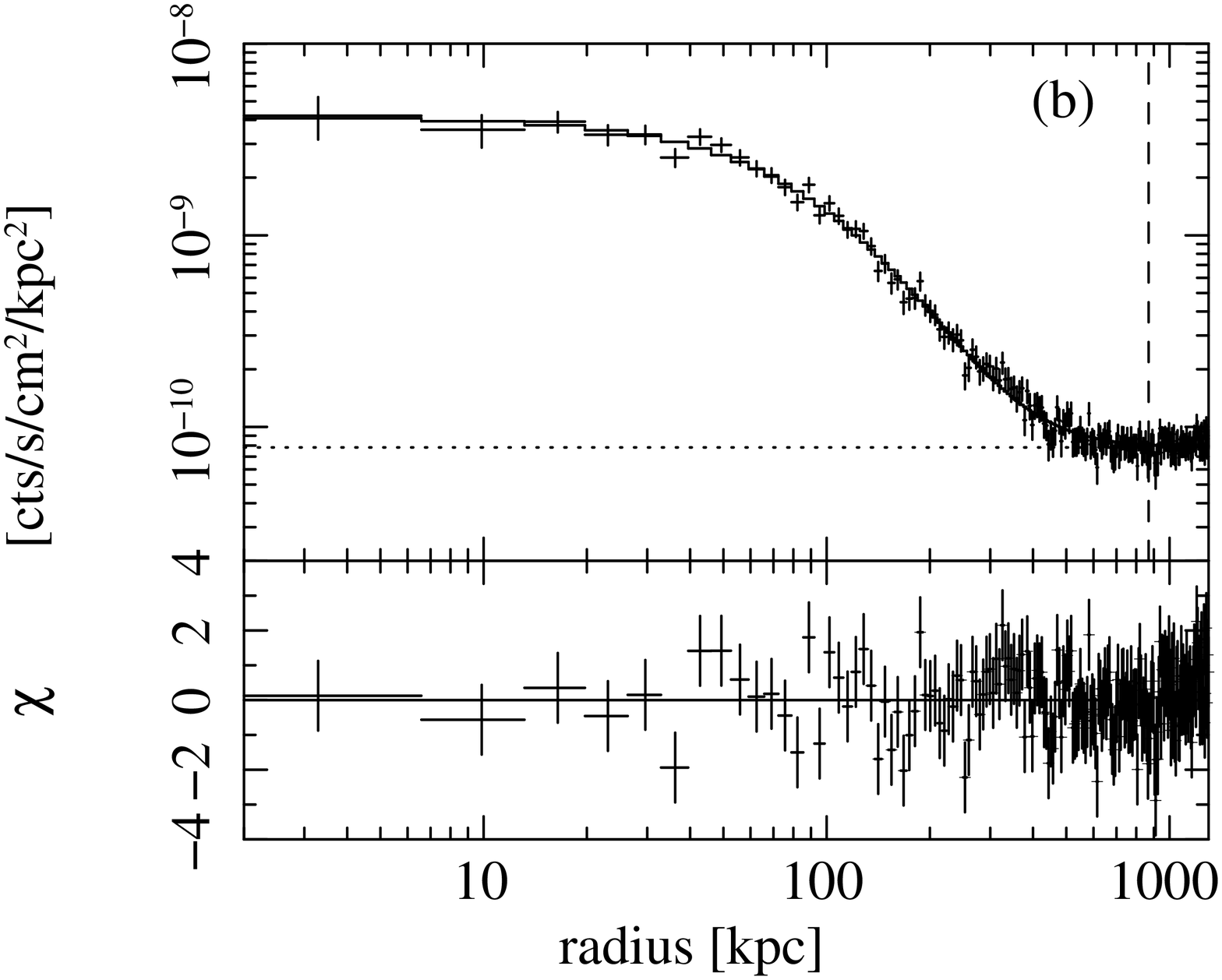}
\caption{Results of the X-ray surface brightness profile fits with
  (a) the $\beta$-model and (b) the NFW-SSM model. In each panel, the
  crosses show the observed surface brightness in the 0.5--5~keV band
  and the solid line shows the best-fit model. The background is shown
  with the horizontal dotted line. The vertical dashed line shows the
  observed extent of the diffuse X-ray emission, $r_{\rm X}$ (see text). 
  The bottom panel shows the residuals in units of the local noise.
\label{fig:sb}}
\end{figure}

We fit the radial profile with the following two models: (1) a
conventional $\beta$-model $S(r)=S_0[1+(r/r_c)^2]^{-3\beta+1/2}$, and
(2) a profile derived from the universal mass profile proposed by
\cite{navarro97} plus isothermality of the cluster \citep[equations
29--32 in ][ hereafter NFW-SSM]{suto98}.  The background level, which
was assumed to be a constant, is simultaneously fit.  As illustrated
in Figure~\ref{fig:sb}, we find that both models can fit the observed
radial profile reasonably well, with reduced chi-square values of
$\chi^2/{\rm dof}=183.2/196$ and $178.9/196$ for the $\beta$-model and
NFW-SSM model, respectively. Note that the extent of the X-ray
emission above the $3\sigma$ background level is derived as $r_{\rm
  X}=130\arcsec (=874)$~kpc.  The fitted values of the $\beta$-model
parameters are $\beta=0.72^{+0.05}_{-0.04}$ and $r_c =
16.4^{+1.8}_{-1.6}~{\rm arcsec} (=109^{+12}_{-11}$~kpc), while the
NFW-SSM model parameters are $B=12.2^{+1.6}_{-1.2}$ and
$r_s=78^{+16}_{-11}~{\rm arcsec}(=519^{+104}_{-77}$~kpc).  Note $B$ is
defined by $B\equiv4\pi G\mu m_p \rho_s r_s^2/kT$ and related to the
outer slope of the radial profile \citep{suto98}.

In the $\beta$-model, the central electron density is $n_{e0} =
(1.5\pm0.1)\times10^{-2}~{\rm cm^{-3}}$.  This implies a radiative
cooling time at the cluster center of $t_{\rm cool} =
5.8^{+1.1}_{-0.9}$~Gyr.  Here $n_{e0}$ and $t_{\rm cool}$ are derived
following basically the same way described in \S5.2.1 of
\citet{ota04b}.  The cooling time is shorter than the Hubble time but
comparable to the age of the Universe at the cluster redshift, $t_{\rm
  age}=8.2$~Gyr. Therefore, radiative cooling should not be important
in the cluster core region.

\begin{figure}
\epsscale{.8}
\plotone{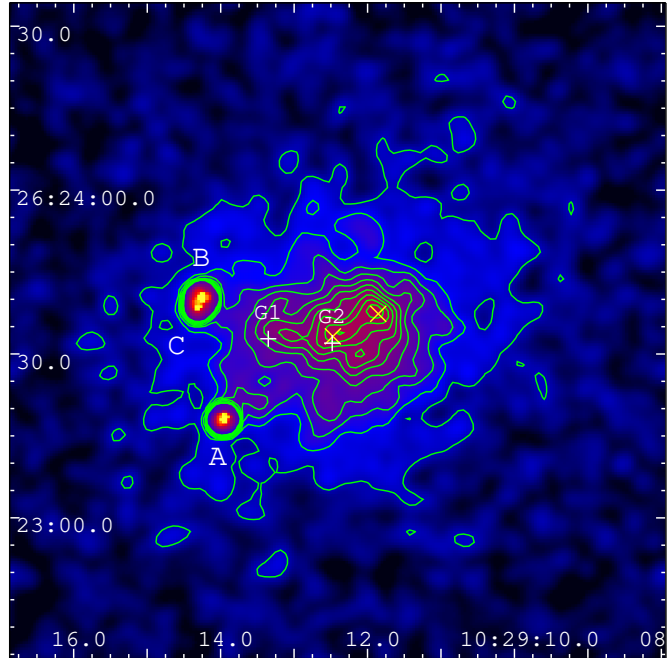}
\caption{Adaptively-smoothed ACIS-S3 image of SDSS~J1029+2623 in the
  0.5--7~keV band.  Both the image and contours are X-ray data.
  The A--C quasar images of SDSS~J1029+2623 and
  the extended emission from the lensing cluster are seen. 
  The X-ray centroid of the cluster component and the North-West
  subpeak are marked with an ``$\times$'' while the positions of the
  two most luminous galaxies, G1 and G2, are marked with a ``$+$''.
  The close B/C image pair is resolved in the raw, unsmoothed image
  (see \S\ref{sec:quasar}). 
\label{fig:image}}
\end{figure}

The {\it Chandra} image suggests that the cluster emission is
elongated along the East-West direction. Moreover, there is another
emission peak about $10\arcsec$ North-West of the cluster centroid
(Fig.~\ref{fig:image}). These motivate us to explore the X-ray surface
brightness with two-dimensional models, adopting both the single- and
double-component elliptical $\beta$-models. The elliptical
$\beta$-model\footnote{beta2d model in the {\tt Sherpa} fitting
  package} is defined by
\begin{eqnarray}
  S(x,y) &=& S_{0} \left[ 1 + \left(\frac{r}{r_{c}}\right)^2 \right]^{-3\beta+1/2} + C, \label{eq:beta2d}\\ 
  r(x,y) &=& \left[ \{ (x - x_{0})\cos{\theta} + (y-y_{0})\sin{\theta}
    \}^2(1-\epsilon)^2 \right.\nonumber\\
  && \left. +\{ (y - y_{0})\cos{\theta} - (x-x_{0})\sin{\theta} \}^2\right]^{1/2} / (1 - \epsilon), \nonumber
\end{eqnarray}
where $r_c$ is the core radius, $x_0$ and $y_0$ define the cluster
center, $\epsilon$ is the ellipticity, $\theta$ is the position angle,
$S_0$ is the amplitude at the center. The background level per image
pixel was fixed to the value obtained from the one-dimensional
analysis, $C=3.3\times10^{-9}~{\rm counts\,s^{-1}\,cm^{-2}}$.

\begin{figure}
\epsscale{.8}
\plotone{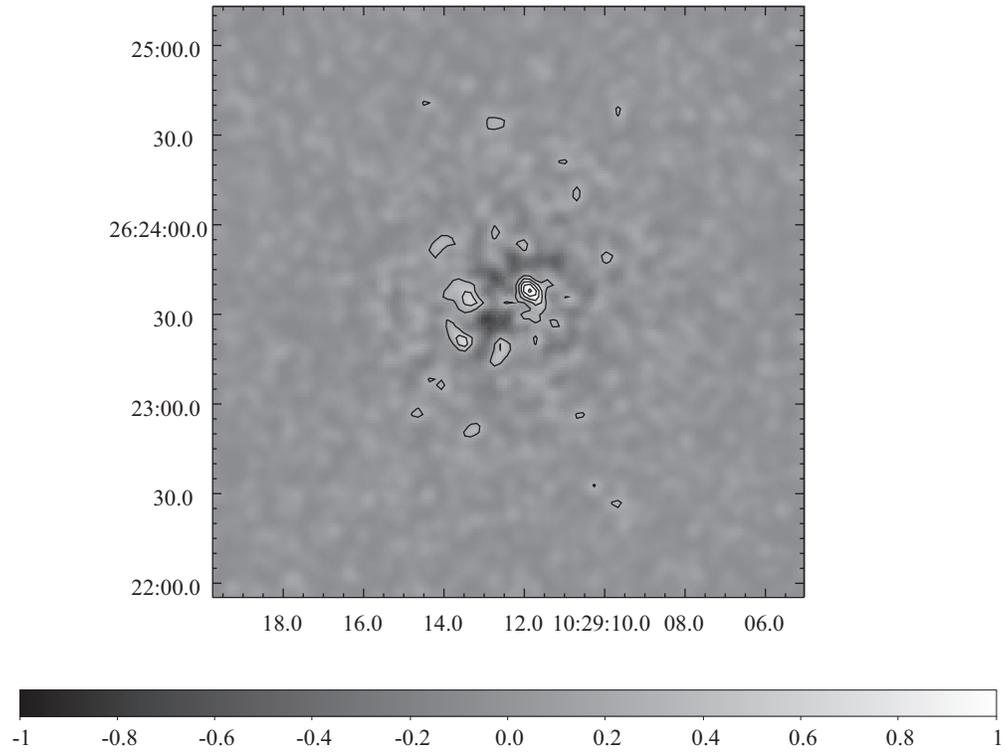}
\caption{Residuals from fits with a single elliptical
  $\beta$-model. The image is smoothed with a $\sigma=2\arcsec$
  Gaussian, and excess  emission over the best-fit model is shown by
  the black contours with linear spacing.  
\label{fig:beta2d_residual}}
\end{figure}

We first fit a single-component elliptical $\beta$-model. The
maximum-likelihood fit was performed with {\tt Sherpa}, where the
exposure map was included to convolve the model image with the
telescope and detector responses.  We find that the fit residuals show
significant excess emission in the North-West region, as shown in
Fig.~\ref{fig:beta2d_residual}. The parameters and results for the
single-component model are summarized in Table~\ref{tab:beta2d}.

\begin{figure}
\epsscale{1.}
\plotone{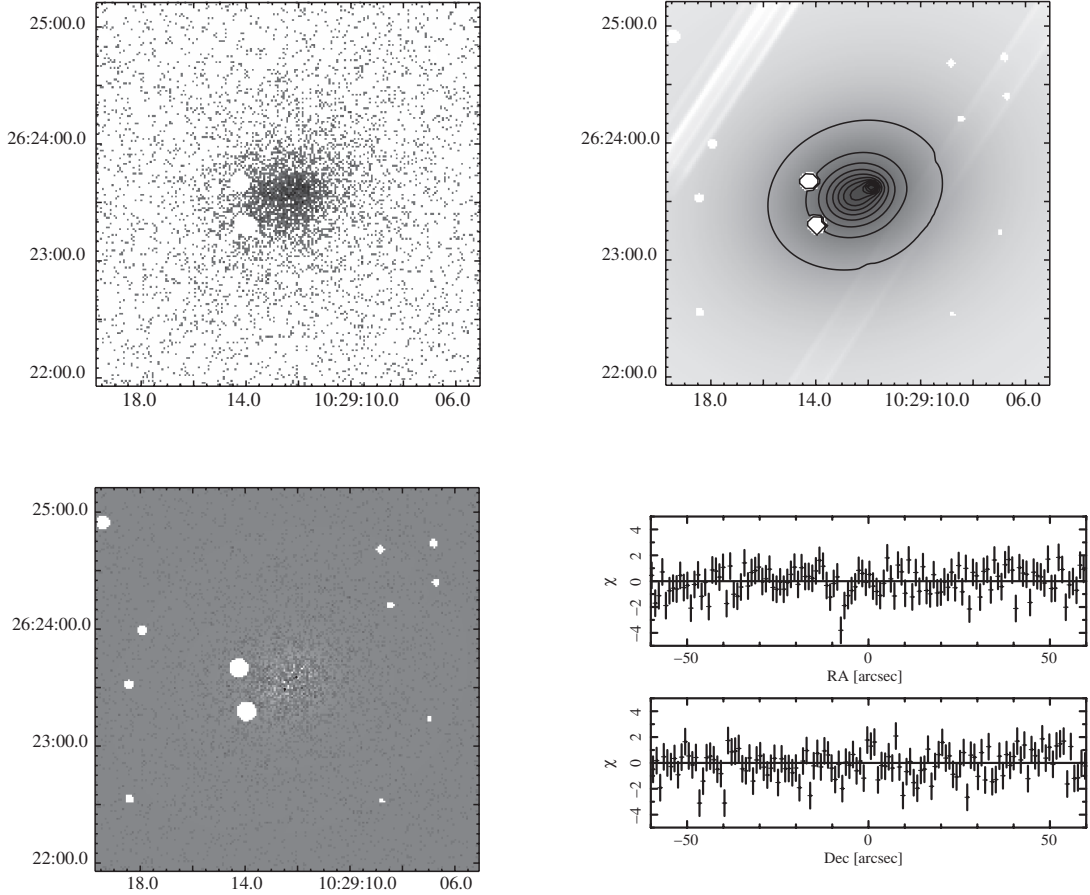}
\caption{Results for the fits with the double elliptical
  $\beta$-model. Top left: {\it Chandra} ACIS-S3 
  image of the central $3\farcm3\times3\farcm3$ region of
  SDSS~J1029+2623 in the 0.5--5 keV energy range. Top right: The
  best-fitting elliptical $\beta$-model, overlaid with logarithmically
  spaced intensity contours. Bottom left: Residuals from the fit.
   Bottom right: Residuals along East-West ({\it top}) and North-South
   ({\it bottom}) lines  through the cluster center in units of
   significance.}
\label{fig:beta2d}
\end{figure}

We next consider a two-component model in order to determine the X-ray
emission profile of the ICM more precisely. The model consists of two
elliptical $\beta$-models, one centered on the cluster and the other
on the North-West component, plus the constant background. For the
North-West component, the parameter $\beta_2$ and position angle
$\theta_2$ were fixed to 0.7 and 0, respectively, due to the limited
photon statistics.  The results from fitting the central
$3\farcm3\times3\farcm3$ region are shown in Table~\ref{tab:beta2d}
and Figure~\ref{fig:beta2d}.  In order to check the goodness of the
fit, we re-binned the image into two single-dimensional profiles in
the Right Ascension and Declination directions (Fig.~\ref{fig:beta2d},
bottom right) and calculated the $\chi^2$ values between the model and
data profiles within the central $\pm 1\arcmin$ region to find that
they are sufficiently small, $\chi^2$/dof = 131/122 and 148/122 for
the $x$- and $y$-directions, respectively. The best-fit cluster-center
position of (10:29:12.47, +26:23:33.2) is consistent with G2
\citep[10:29:12.48, +26:23:32.0;][]{inada06} within the errors. The
ellipticity of the main cluster component is measured to be
$\epsilon=0.214^{+0.034}_{-0.035}$ (axis ratio $q=0.786$), which is
typical of X-ray surface brightness ellipticities
\citep{defilippis05,flores07,hashimoto07,kawahara10,lau12}.

Note that there appears to be no optical counterpart for the NW peak
at (10:29:11.9, +26:23:37.4).  For the APEC model, the X-ray spectrum
extracted from $r<5\arcsec$ around the NW peak with subtracted
background from the surrounding $5\arcsec<r<10\arcsec$ region yields
$kT= 3.8(>1.7)$~keV and an unabsorbed bolometric luminosity of
$1.5\times10^{43}~{\rm erg\,s^{-1}}$.  These parameters are
reminiscent of the emission peak detected in the line-of-sight merging
system \object{CL0024+17} \citep{ota04a}. In addition, there is a
possible jump in the X-ray surface brightness distribution near the NW
peak (Fig.~\ref{fig:image}). This might be the signature that the NW
component is in the process of merging with the main cluster or has
recently undergone a merger.  These observed features suggest that the
NW peak may be a remnant of a merging core or a contact discontinuity
so-called cold front \citep{markevitch00,owers09}, although the
limited photon statistics\footnote{The source and background are
  estimated to be 95 and 163~counts within $r<5\arcsec$,
  respectively.} prevent a more detailed exploration.

\subsection{Cluster mass distribution}\label{subsec:cluster_mass}

\begin{figure}
\epsscale{.8}
\plotone{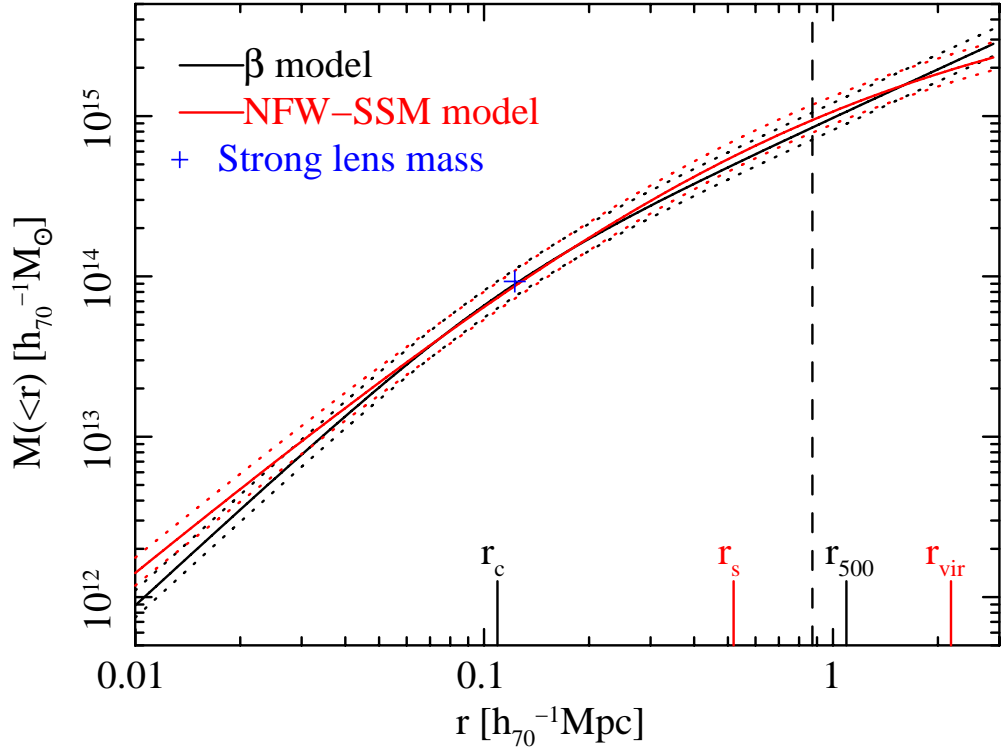}
\caption{Enclosed mass of the lensing cluster, $M_{\rm X}$, for the
  $\beta$-model (black) and the NFW-SSM model (red).  The dotted lines
  indicate the 90\% error ranges.  Note that $M _{\rm X}$ is a
  cylindrical cluster mass projected within a radius $r$. The mass
  derived from gravitational lensing is also shown for comparison. 
  The meaning of the vertical dashed line is the same as Figure~\ref{fig:sb}.
\label{fig:mx}}
\end{figure}

Under the assumption of hydrostatic equilibrium and spherical
symmetry, we can infer the mass distribution of the lensing cluster
from the X-ray temperature and surface-brightness profiles.  We assume
isothermality because we did not detect a significant radial
dependence of the X-ray temperature. Figure~\ref{fig:mx} shows the
cylindrical cluster mass projected within a radius $r$ derived from
each surface mass distribution profile. We find that both models yield
consistent mass profiles within the measurement errors.

For the NFW-SSM model and the critical overdensity
$\Delta_c=\Delta_{\rm vir}=18\pi^2\Omega^{0.427}$ \citep{nakamura97},
the virial mass and the concentration parameter (defined by the ratio
of the virial radius to the scale radius of the NFW profile) are
constrained to be $M_{\rm vir}=1.57^{+0.66}_{-0.39}\times
10^{15}M_\odot$ and $c_{\rm vir}=4.52^{+0.58}_{-0.58}$. However, these
values should be taken with caution because they rely on the
extrapolation of the profiles well beyond the radii where X-ray
properties are measured. For comparison with other mass measurements,
it may be better to use masses defined for larger critical
overdensities $\Delta_c$. In Table~\ref{tab:mass} we summarize the
mass and radius estimated for the virial overdensity and overdensities
of $\Delta_c=500$, $1000$, and $2500$, for both the $\beta$-model and
the NFW-SSM model.  The result indicates that only $r_{2500}$ is
smaller than the extent of the X-ray emission, $r_{\rm X}$, presented
in \S\ref{subsec:cluster_image}. This means that the results for the
other overdensities rely on the extrapolation of the data.

Next we compare the mass distribution derived from the X-ray analysis
with those from strong lens models. An issue for strong lens modeling
for this cluster has been the uncertainty of the mass centroid,
because the naked cusp configuration poorly constrains the center of
the mass distribution \citep{inada06,oguri08}. Given that the
measurement of the X-ray centroid agrees well with the location of one
of the brightest cluster galaxies (galaxy G2), it is reasonable to
assume that the centroid coincides with galaxy G2, because if the
cluster gas is significantly segregated from the dark matter
distribution due to merging \citep[e.g.,][]{markevitch04,bradac08}
then there should be no correlation between X-ray centroids and bright
galaxies that trace the dark matter distribution. We fit the positions
of the three quasar images assuming an NFW profile plus external shear
model using the software {\it glafic} \citep{oguri10}. We find that
the best-fit model predicts the Einstein radius of $\theta_{\rm
  E}=18\farcs3$ for the source redshift of $z_s=2.197$. This
translates into a cylindrical cluster mass of $M_{\rm 2D}(<122{\rm
  kpc})=9.30\times10^{13}M_\odot$, which is in good agreement with the
enclosed mass derived from the X-ray data (see
Figure~\ref{fig:mx}). The rough agreement between lensing and X-ray
masses implies that the effect of the line-of-sight elongation of the
cluster mass distribution is not significant, despite the fact that
possible merging along the line-of-sight suggests a large projection
effect.  On the other hand, at larger radii ($r \sim 0.5$~Mpc), the
X-ray mass is found to be significantly larger than the lensing mass
derived from a recent lensing analysis of {\it Hubble Space Telescope}
(HST) data (Oguri et al. in prep.). The mass discrepancy is ascribed
to a temperature enhancement due to shock heating during a merger,
which might lead to the overestimate of X-ray mass near the Einstein
radius. See Oguri et al. in prep. for further discussion.

\section{Quasar images}\label{sec:quasar}

\subsection{X-ray properties}\label{subsec:qso_flux}

\begin{figure}
\plotone{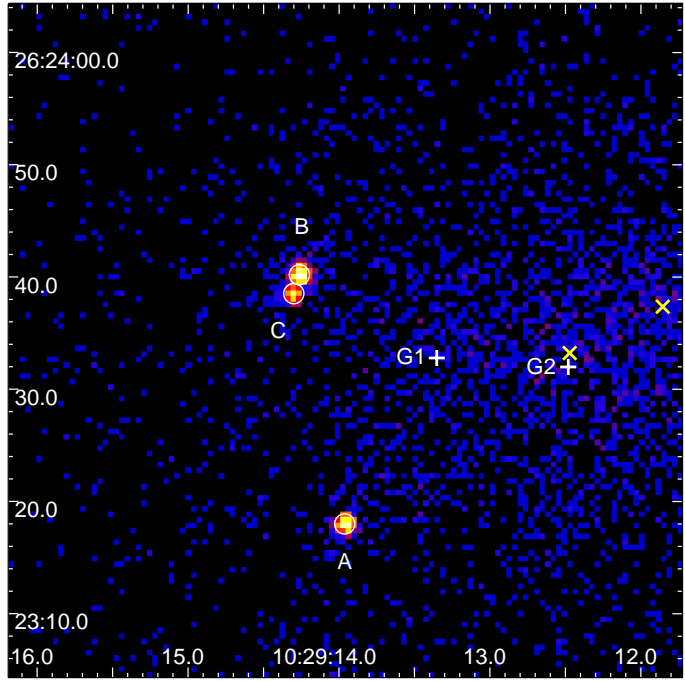}
\caption{Raw ACIS-S3 image of  SDSS~J1029+2623 in the 0.5--7~keV band. The spectral regions for quasar images A--C are indicated with the circles. 
The meaning of the labels are the same as Figure~\ref{fig:image}.
\label{fig:rawimage}}
\end{figure}

Given the small $1\farcs8$ image separation of images B and C (see
Figure~\ref{fig:rawimage}), we measure the X-ray fluxes and spectra of
the quasar images using a circular aperture with a radius of
$0\farcs9$ for all the quasar images. The observed counts in the
0.5--7~keV band are 687, 1073, and 295, for images A--C,
respectively. The net contamination from the background and the
cluster is estimated to be only a few percent. Similarly the
contamination from image B in the extraction radius for image C is
estimated to be $\sim 10$~counts, which is smaller than the $1\sigma$
statistical error of the flux of image C. We do not find any clear
sign of time variability for light curves produced with time
resolutions of 2048 or 4096~s.

\begin{figure}
\epsscale{1.1}
\plottwo{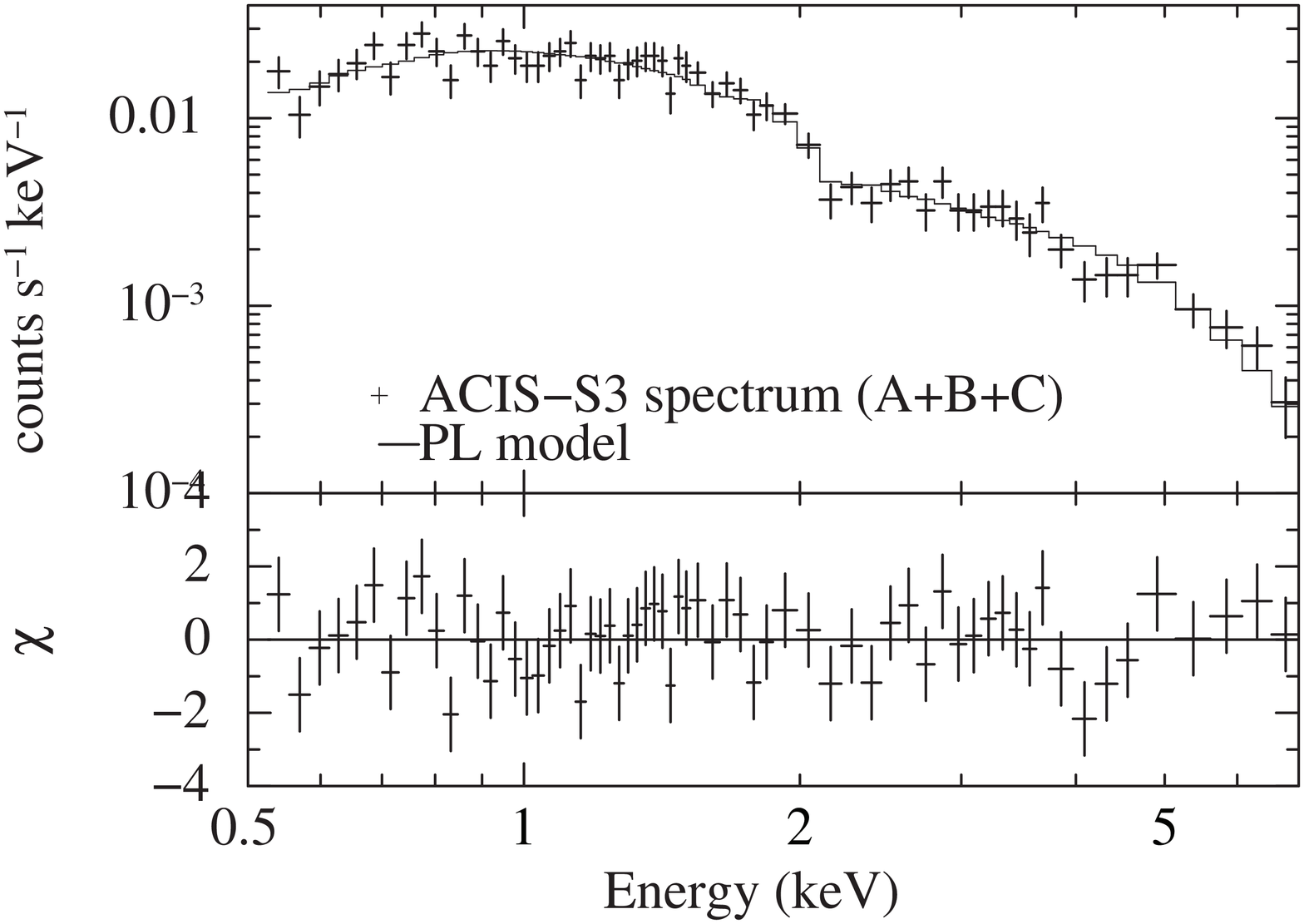}{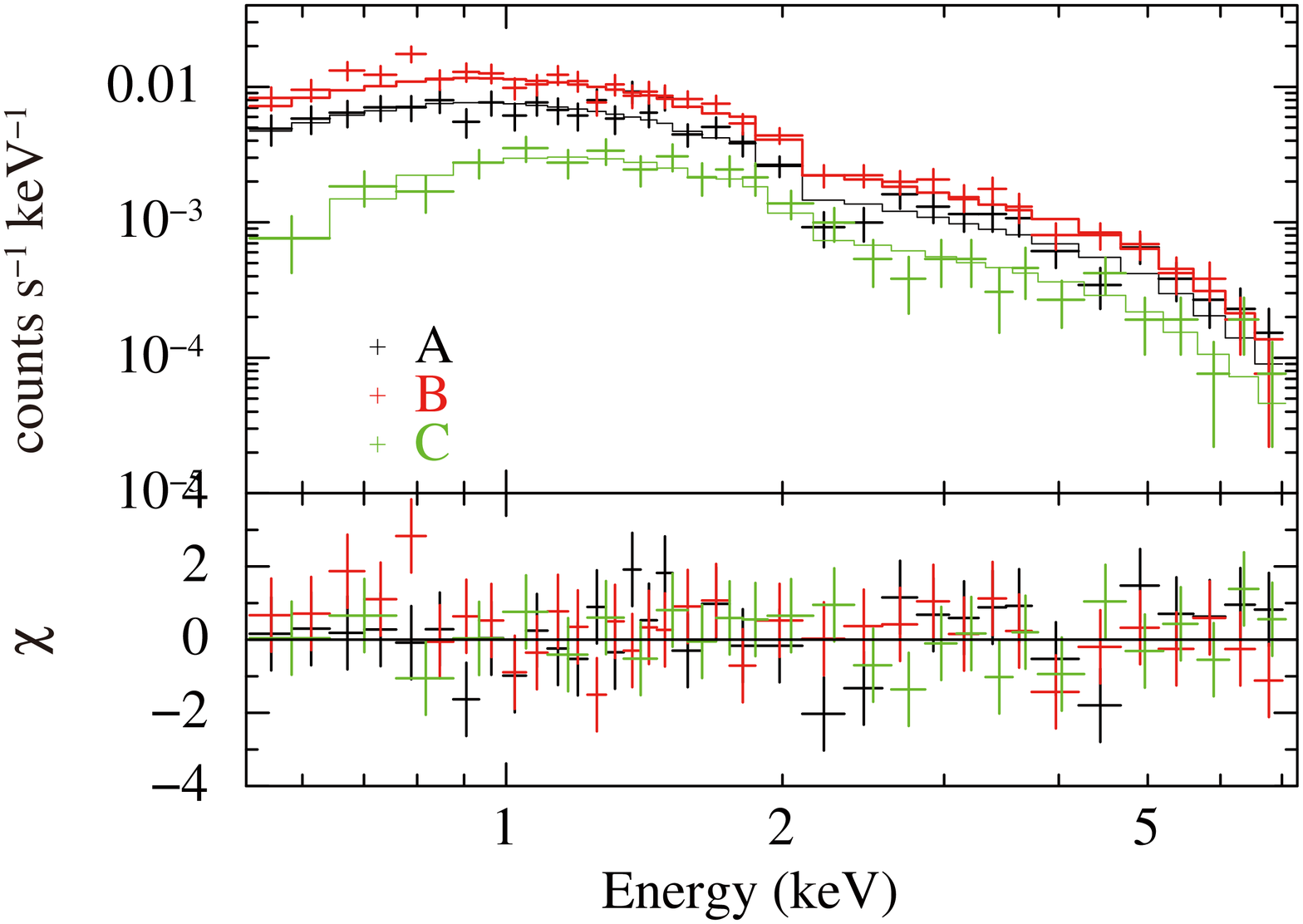}
\caption{Left: total quasar spectrum fit with the power-law model (solid line). Right: Spectra of lensed quasar images A (black), B (red), and C
(green) fit with a common intrinsic slope $\Gamma$. In each top panel, the crosses denote the spectrum in the observed frame.
\label{fig:qso_spec}}
\end{figure}

The combined spectrum of the three lensed images A, B, and C is shown
in the left panel of Figure~\ref{fig:qso_spec}. We fit the spectrum
with a power-law model.  The Galactic absorption is again fixed at
$N_{\rm H}=1.67\times 10^{20}{\rm cm^{-2}}$. We do not find
significant intrinsic absorption in the total spectrum. The power-law
model is acceptable at the 90\% confidence level ($\chi^2/{\rm
  dof}=57.1/63$) and the power-law index of $\Gamma=1.55\pm0.06$
agrees well with the mean value for mini-BAL quasars observed with
{\it Chandra} \citep{gibson09}. The corresponding total luminosity is
$5.3\times10^{45}~{\rm ergs\,s^{-1}}$ in the 2--10~keV band.  We do
not detect any significant emission from the neutral iron K$\alpha$
line at 2.0~keV in the observed frame (or 6.4~keV in the rest
frame). The 90\% upper limit on the equivalent width is ${\rm EW} <
170 $~eV in the quasar rest frame for a power-law combined with a
Gaussian line model at 2~keV.

Next we study the spectra of the individual quasar images. We first
check the hardness ratios defined by the count rate ratio in the
2--7~keV and 0.5--2~keV bands.  The spectral shape of image C differs
from A and B because the hardness ratio for image C ($0.48\pm0.06$)
differs from those of images A ($0.37\pm0.03$) and B
($0.36\pm0.02$). Indeed, the spectra of images A--C plotted in the
right panel of Figure~\ref{fig:qso_spec} clearly indicate that the
emission of image C is heavily absorbed in the soft band. To quantify
this, we fit the spectrum with a power-law model absorbed by
intervening cold material in addition to the Galactic interstellar
medium. Since the location of the cold absorber is not known, we
consider cases with the absorber located at either the quasar redshift
of $z_s=2.197$ or at the lens redshift of $z_l=0.584$.

The result of the power-law plus the absorption model fits are
summarized in Table~\ref{tab:qso_spec}.  While the inclusion of the
additional absorption does not improve the best-fit $\chi^2$ value for
images A and B, the fits to image C are significantly improved by
including the absorption. We find that the improvement is significant
at the $>99.98$\% confidence level according to the F-test ($\Delta
\chi^2 =11$ for one additional parameter), although the redshift of
the absorber is not constrained.  While the power-law model without
the additional absorption for image C appears to be acceptable in the
sense that the reduced $\chi^2$ is very close to 1, the resulting
power-law slope of $\Gamma=1.21$ is siginificantly smaller compared
with the best-fit slope of $\Gamma\sim 1.6$ for images A and B. When
$\Gamma$ is fixed to 1.55, which is the best-fit slope for the total
quasar spectrum, the reduced $\chi^2$ for image C is
$\chi^2$/dof=39.5/26 without the additional absorption, indicating
that the fit is not acceptable at 90\% confidence level.  We also
tried a simultaneous fit to the three spectra with a common intrinsic
spectral index $\Gamma$, because the intrinsic photon indices are
statistically consistent among the three images. The result shown in
Table~\ref{tab:qso_spec_2} again indicates that the extra absorption
is significant only for image C.

\subsection{Comparison with optical and radio flux ratios}

In Table~\ref{tab:qso_flux}, we compare flux ratios between quasar
images measured in the {\it Chandra} X-ray images with radio and
optical flux ratios presented in \citet{kratzer11}.  We find that the
absorption corrected flux ratio $C/B$ in the X-ray is larger than the
optical flux ratios and is close to the radio flux ratio.

The origin of the absorption may be inferred from the dust-to-gas
ratio. Based on the optical spectra of images B and C in
\citet{oguri08}, we estimated a color excess of $E(B-V)\simeq 0.07$
for absorption at $z_s=2.197$ and $E(B-V)\simeq 0.17$ for absorption
at $z_l=0.584$. These imply dust-to-gas ratios of $E(B-V)/N_H\simeq
2\times 10^{-24}{\rm mag}\,{\rm cm^2}{\rm atom^{-1}}$ and
$E(B-V)/N_H\simeq 3\times 10^{-23}{\rm mag}\,{\rm cm^2}{\rm
  atom^{-1}}$ at the source and lens redshifts, respectively. The
higher value found for dust in the lens is broadly consistent with
measurements for other lens systems \citep{dai09}, while the value for
dust at the source redshift is much lower than expected from this
observed correlation.  However, such low dust-to-gas ratios appear to
be common for intrinsic absorption features of quasars caused by
quasar outflow \citep{hall06}, which can be explained by dust
evaporation due to irradiation by the central source.  Thus we cannot
draw any firm conclusion about the origin of the absorption from this
analysis. Both the mini-BAL features in the quasar spectrum
\citep[see][]{oguri08}, which assures the present of outflows in the
sightline of this quasar, and the deficiency of dust in the
intracluster medium particularly near the center
\citep{chelouche07,muller08,bovy08,kitayama09} suggest that intrinsic
absorption at the quasar redshift is the more likely scenario,
although the anomalous flux ratios imply a faint galaxy near image C
which might host the dust at the lens redshift. Indeed, a probable
galaxy has recently been identified in the vicinity of image C in the
HST image (Oguri et al. in prep.), which might cause the extra
absorption detected in image C. If the former interpretation is
correct, the different absorptions between image B and C probe spatial
variations of the absorption content, possibly related to the
structure of disk wind outflows \citep{green06}.

We note that the optical $C/B$ flux ratio becomes $C/B\sim 0.4$ once
the dust extinction is corrected (see below), which is also consistent
with the radio and absorption-corrected X-ray flux ratios. The nearly
consistent flux ratios between the radio, optical, and X-ray
observations imply that microlensing effect is not significant for
this system, particularly because X-ray flux ratios are more easily
affected by microlensing \citep[e.g.,][]{ota06b,pooley07,dai10}.

\section{Summary and Discussions}
We have presented results from {\it Chandra} observations of
SDSS~J1029+2623, the largest-separation gravitational lens system,
produced by a cluster of galaxies at $z=0.58$.  We detect significant
X-ray emission from the lensing cluster as well as three lensed quasar
images.

From the extended cluster emission, the X-ray temperature and
bolometric luminosity are constrained to be $kT=8.1^{+2.0}_{-1.2}$~keV
and $L_{\rm X} = 9.6\times10^{44}~{\rm erg\,s^{-1}}$.  The luminosity
is lower than the mean value expected from the luminosity-temperature
relation of distant clusters, but it is within the observed scatter of
the relation.  While the estimated cooling time is shorter than the
Hubble time, the lack of a significant temperature drop at the cluster
center suggests that radiative cooling is not important in this
system. From the image analysis, we have found that the cluster
emission is elongated along the East-West direction and has a subpeak
in the North-West region. This indicates that the system has undergone
a merger. However, the modest ellipticity of the gas distribution
found in the main cluster component indicates that the irregularity of
mass distribution may not be large. We have reconstructed the mass
profile of the lensing cluster, assuming isothermality and hydrostatic
equilibrium to find that the X-ray mass within the Einstein radius or
$\sim 120$~kpc is in a good agreement with that expected from the
strong lensing. The agreement seems to imply that there is not any
significant departure from the equilibrium state in the cluster core,
and that the line-of-sight elongation of this cluster may not be so
large. On the other hand, the mass discrepancy between the X-ray and
lensing measurements was found at larger radii, indicating a complex
nature of the mass distribution of this system (Oguri et al. in
prep.).

The X-ray spectra of the three quasar images above 2~keV are well
represented by a power-law with a photon index of $\Gamma\sim1.55$
that is typical of mini-BAL quasars. However, significant soft X-ray
absorption due to the cold material is detected only in image C. After
correcting for the X-ray absorption, the X-ray flux ratios between the
quasar images are roughly consistent with those in optical and radio,
which suggests that microlensing is not significant for this
system. This indicates that the flux ratio anomaly seen in this lens
system has to be due to substructure \citep[see, e.g.,][]{kochanek04}.
It is difficult to distinguish the possibilities between absorbers at
the quasar redshift and at the cluster redshifts from the X-ray
spectrum or the dust-to-gas ratio.

Both the surface brightness profile of the lensing cluster and the
anomalous flux ratio are better understood by comparing the result
with accurate mass modeling results obtained from gravitational
lensing analysis. For this purpose we have recently obtained {\it HST}
images of this system. Details of lensing analysis results of the {\it
  HST} data and a comparison with the {\it Chandra} results will be
presented in a forthcoming paper (Oguri et al. in prep.).

\acknowledgments

We thank K. Aoki for useful discussions, and an anonymous referee for
many useful suggestions.  This work was supported in part by the
Grant-in-Aid for Scientific Research by the Ministry of Education,
Culture, Sports, Science and Technology, No.22740124 (N.O.).  This
work was supported in part by the FIRST program ``Subaru Measurements
of Images and Redshifts (SuMIRe)'', World Premier International
Research Center Initiative (WPI Initiative), MEXT, Japan, and
Grant-in-Aid for Scientific Research from the JSPS (23740161).
C.S.K. is supported by NSF grant AST-1009756.  X.D. acknowledges
support by NASA/SAO fund GO0-11147B.  T.S. acknowledges support from
NSF through grant AST-0444059-001 and the Smithsonian Astrophysics
Observatory through grant GO0-11147A.

{\it Facilities:} \facility{CXO (ACIS)}.

\clearpage

\begin{table}
\begin{center}
\caption{APEC Models of the Overall Spectrum}
\label{tab:cluster_spec}
\begin{tabular}{ll}\\\hline\hline
Parameter  & Value (90\% error) \\\hline
$N_{\rm H}$ (${\rm cm^{-2}}$) & $1.67\times10^{20}$ (F) \\
$kT$ (keV)  &   8.1 (6.8 -- 10.1)  \\
Abundance ($Z_{\sun}$) & $0.22 (< 0.44)$ \\
Redshift & 0.584 (F) \\
Normalization\tablenotemark{a} & 7.31 (6.93 -- 7.68)$\times10^{-4}$  \\
$\chi^2/dof $ &47.5/45 \\
$f_{{\rm X}, 0.5-7}$ (${\rm erg\,s^{-1}\,cm^{-2}}$)\tablenotemark{b} & $4.6\times10^{-13}$ \\
$L_{{\rm X}, 0.5-7}$ (${\rm erg\,s^{-1}}$)     \tablenotemark{c}& $5.5\times 10^{44}$ \\
$L_{{\rm X}, {\rm bol}}$ (${\rm erg\,s^{-1}}$) \tablenotemark{d}& $9.6\times 10^{44}$ \\\hline
\end{tabular}
\end{center}
\tablenotetext{a}{Normalization factor for the APEC model, $\int n_e
n_{\rm H} dV/4\pi D_A^2 (1+z)^2$ in $10^{-14}{\rm cm^{-5}}$, 
where $D_A$ is angular size distance to the cluster.}  
\tablenotetext{b}{Galactic absorption-corrected X-ray flux in the 0.5--7 keV band within $r<1\arcmin$.}
\tablenotetext{c}{Galactic absorption-corrected X-ray luminosity in the 0.5--7 keV band within $r<1\arcmin$.}
\tablenotetext{d}{Bolometric luminosity within $r_{500}$.}
\tablenotetext{}{(F) Fixed parameters.} 

\end{table}

\begin{deluxetable}{ccc}
\tablewidth{0pt}
\tablecaption{Results for Circularly Symmetric  Models\label{tab:beta1d}}
\tablehead{\colhead{Model}&\colhead{Parameter} & \colhead{Value}}
\startdata
$\beta$-model & $S_{0}$~(${\rm counts\,s^{-1}\,cm^{-2}\,kpc^{-2}}$)\tablenotemark{a} & $3.57^{+0.28}_{-0.26}\times10^{-9}$ \\
 & $\beta$        & $0.72^{+0.05}_{-0.04}$\\
 & $r_c$~(arcsec/kpc) & $16.4^{+1.8}_{-1.6}/109^{+12}_{-11}$\\ 
 & $\chi^2/dof$ & 183.2/196\\ \hline
NFW-SSM model & $S_{0}$~(${\rm counts\,s^{-1}\,cm^{-2}\,kpc^{-2}}$)\tablenotemark{a}& $4.03^{+0.31}_{-0.39}\times10^{-9}$\\
 & $B$        & $12.2^{+1.6}_{-1.2}$\\
 & $r_s$~(arcsec/kpc) & $78^{+16}_{-11}/519^{+104}_{-77}$\\ 
 & $\chi^2/dof$ & 178.9/196\\ 
\enddata
\end{deluxetable}

\begin{deluxetable}{ccc}
\tablewidth{0pt}
\tablecaption{Results for Elliptical Models\label{tab:beta2d}}
\tablehead{\colhead{Model}&\colhead{Parameter} & \colhead{Value}}
\startdata
Single elliptical $\beta$-model & $x_{0,1}, y_{0,1}$ & 10:29:12.395, +26:23:33.71\tablenotemark{a}\\
& $S_0$~(${\rm counts\,s^{-1}\,cm^{-2}\,kpc^{-2}}$) & $3.57^{+0.28}_{-0.25}\times10^{-9}$ \\ 
& $\beta$ & $0.70^{+0.04}_{-0.03}$ \\
& $r_c$~(arcsec/kpc) & $17.6^{+1.7}_{-1.6}$/$118^{+12}_{-10}$\\
& $\epsilon$ & $0.214^{+0.034}_{-0.035}$  \\
& $\theta$ & $0.36^{+0.09}_{-0.09}$ \\ \hline
Double elliptical $\beta$-model & $x_{0,1}, y_{0,1}$ & 10:29:12.471, +26:23:33.24\tablenotemark{b}\\
& $S_{0,1}$~(${\rm counts\,s^{-1}\,cm^{-2}\,kpc^{-2}}$) & $3.15^{+0.36}_{-0.61}\times10^{-9}$ \\ 
& $\beta_1$ & $0.71^{+0.04}_{-0.04}$ \\
& $r_{c,1}$~(arcsec/kpc) & $19.0^{+2.4}_{-1.9}$/$127^{+16}_{-13}$\\
& $\epsilon_{1}$ & $0.214^{+0.034}_{-0.035}$  \\
& $\theta_1$ & $0.36^{+0.09}_{-0.09}$ \\ 
& $x_{0,2}, y_{0,2}$ & 10:29:11.853, +26:23:37.37\tablenotemark{c}\\
& $S_{0,2}$~(${\rm counts\,s^{-1}\,cm^{-2}\,kpc^{-2}}$) & $3.1^{+2.3}_{-1.2}\times10^{-9}$ \\ 
& $\beta_2$ & 0.70(F) \\
& $r_{c,2}$~(arcsec/kpc) & $3.9^{+4.1}_{-1.7}$/$26^{+28}_{-12}$\\
& $\epsilon_{2}$ & $0.19(<0.35)$  \\
& $\theta_2$ & 0.0(F) 
\enddata
\tablenotetext{a}{The 90\% errors are $\pm1\farcs1$ for $x_{0}$ and $\pm1\farcs0$ for $y_{0}$.}
\tablenotetext{b}{The 90\% errors are $\pm1\farcs4$ for $x_{0,1}$ and $\pm1\farcs1$ for $y_{0,1}$.}
\tablenotetext{c}{The 90\% errors are $\pm1\farcs7$ for $x_{0,2}$ and $\pm1\farcs5$ for $y_{0,2}$.}
\end{deluxetable}

\begin{center}
\begin{table}
\caption{Masses for Different Critical Overdensities}\label{tab:mass}
\begin{tabular}{ccccccccc}\hline\hline
Model & $M_{\rm vir}$\tablenotemark{a} &  $r_{\rm vir}$\tablenotemark{b} & 
$M_{500}$\tablenotemark{a} &  $r_{500}$\tablenotemark{b} & 
$M_{1000}$\tablenotemark{a} &  $r_{1000}$\tablenotemark{b} & 
$M_{2500}$\tablenotemark{a} &  $r_{2500}$\tablenotemark{b} \\ \hline
$\beta$ & $12.95^{+5.00}_{-3.07}$ & $2.10^{+0.24}_{-0.18}$
        & $ 6.68^{+2.60}_{-1.59}$ & $1.09^{+0.13}_{-0.09}$
        & $ 4.65^{+1.83}_{-1.11}$ & $0.77^{+0.09}_{-0.07}$
        & $ 2.81^{+1.13}_{-0.68}$ & $0.48^{+0.06}_{-0.04}$ \\
NFW-SSM & $15.67^{+6.55}_{-3.93}$ & $2.24^{+0.28}_{-0.21}$
        & $ 9.28^{+3.93}_{-2.41}$ & $1.22^{+0.15}_{-0.12}$
        & $ 6.45^{+2.81}_{-1.72}$ & $0.86^{+0.11}_{-0.08}$
        & $ 3.52^{+1.62}_{-0.99}$ & $0.52^{+0.07}_{-0.05}$ \\ \hline
\tablenotetext{a}{In units of $10^{14}M_{\odot}$.}
\tablenotetext{b}{In units of Mpc.}
\end{tabular}
\end{table}
\end{center}

\begin{center}
\begin{table}
\caption{Absorbed Power-law Models for the Quasar Images\label{tab:qso_spec}}
\begin{tabular}{ccccccccc}\hline\hline
          & \multicolumn{2}{c}{No extra absorption}   & \multicolumn{3}{c}{Absorption at $z=2.197$} & \multicolumn{3}{c}{Absorption at $z=0.584$} \\ \cline{2-3} \cline{4-6} \cline{7-9}
Name & $\Gamma$ & $\chi^2/{\rm dof}$ 
& $\Gamma$ & $N_{\rm H}$\tablenotemark{a}  & $\chi^2/{\rm dof}$ 
& $\Gamma$ & $N_{\rm H}$\tablenotemark{a}  & $\chi^2/{\rm dof}$ \\ \hline
A & $1.56^{+ 0.12}_{-0.11}$ & $31.3/33 $  & $1.56^{+ 0.16}_{-0.11}$ & $0.00(<0.70) $ & $31.3/32 $ & $1.56^{+0.16}_{ -0.11}$ & $0.00(< 0.12)$ & $31.3/32$ \\
B & $1.68^{+ 0.09}_{-0.09}$ & $23.2/33 $ & $1.68^{+0.10}_{-0.09}$ & $0.00(<0.33)$ & $23.2/32$ & $1.68^{+0.10}_{  -0.09}$ & $0.00(<0.06)$ & $23.2/32$\\
C & $1.21^{+ 0.14}_{-0.14}$ & $25.1/25 $ & $1.68^{+0.32}_{-0.28}$ & $3.13^{+2.24}_{-1.68}$ & $13.5/24$ & $1.71^{+0.34}_{ -0.30}$ & $0.54^{+0.40}_{-0.30}$ & $14.1/24$\\ \hline
Total & $1.55^{+ 0.06}_{-0.06}$ & $57.1/63 $ & $1.55^{+0.10}_{-0.06}$ & $0.00(<0.46)$ & $57.1/62$ & $1.55^{+0.10}_{ -0.06}$ & $0.00(<0.08)$ & $57.1/62$\\\hline
\end{tabular}
\tablenotetext{a}{The hydrogen column density in the unit of $~{\rm [10^{22}~cm^{-2}]}$.}
\end{table}
\end{center}

\begin{center}
\begin{table}
\caption{Joint fits to the Quasar Spectra}\label{tab:qso_spec_2}
\begin{tabular}{ccccccc}\hline\hline
          & \multicolumn{3}{c}{Absorption at $z=2.197$}           & \multicolumn{3}{c}{Absorption at $z=0.584$} \\ \cline{2-4} \cline{5-7}
Name & $\Gamma$ & $N_{\rm H}$\tablenotemark{a} & $\chi^2/{\rm dof}$ 
& $\Gamma$ & $N_{\rm H}$\tablenotemark{a} & $\chi^2/{\rm dof}$ \\ \hline
A & $1.65^{+ 0.08}_{-0.08}$ & $0.24(<0.89) $ & $69.4/90 $ & $1.65^{+0.08}_{ -0.08}$ & $0.04(< 0.15)$ & $70.0/90$ \\
B &  & $0.00(<0.25)$ &  &  & $0.00(<0.04)$ & \\
C &  & $2.97^{+1.38}_{-1.08}$ &  &  & $0.49^{+0.22}_{-0.18}$ & \\ 
\hline
\end{tabular}
\tablenotetext{a}{The hydrogen column density in the unit of $~{\rm [10^{22}~cm^{-2}]}$.}
\end{table}
\end{center}

\begin{deluxetable}{cccccccc}
\tablewidth{0pt}
\tabletypesize{\footnotesize}
\tablecaption{X-ray and optical properties of SDSS~J1029+2623\label{tab:qso_flux}}
\tablehead{\colhead{Name}
& \colhead{$F_{\rm X}$\tablenotemark{a,b}}
& \colhead{hardness\tablenotemark{c}}
& \colhead{$\alpha_{\rm ox}$}
& \colhead{$f({\rm X})$\tablenotemark{b,d}}
& \colhead{$f({\rm opt}, V)$\tablenotemark{d}}
& \colhead{$f({\rm opt}, I)$\tablenotemark{d}}
& \colhead{$f({\rm radio})$\tablenotemark{d}}
}
\startdata
A & $0.90\pm0.03$ (0.89) & $0.37\pm0.03$ & $-1.32$ 
& 0.66 (0.64) & 0.96 & 0.95& $0.80\pm 0.06$ \\ 
B & $1.37\pm0.04$ (1.34) & $0.36\pm0.02$ & $-1.26$ 
& $\equiv1$  & $\equiv1$ & $\equiv1$  & $\equiv1$  \\ 
C & $0.49\pm0.03$ (0.40) & $0.48\pm0.06$ & $-1.14$ 
& 0.36 (0.27) & 0.16\tablenotemark{e} & 0.24\tablenotemark{e} & $0.46\pm0.05$ \\ 
\enddata
\tablenotetext{a}{Absorption-corrected, 0.5--7~keV X-ray flux in units of 
  $10^{-13}{\rm erg\, s^{-1}cm^{-2}}$ and the 1-$\sigma$ error.}
\tablenotetext{b}{The value in the parenthesis are those without
  absorption corrections.}
\tablenotetext{c}{Hardness ratios defined by the count rate ratios in
  $2-7$~keV and $0.5-2$~keV bands. Errors indicate 68\% confidence limits.}
\tablenotetext{d}{Flux normalized by the flux of image B. The optical and
  radio flux ratios are taken from \citet{kratzer11}. Statistical
  errors on the optical flux ratios are $\la 2\%$ \citep{oguri08}. }  
\tablenotetext{e}{The optical $C/B$ flux ratio becomes $C/B\sim 0.4$
  once the dust extinction is corrected.} 
\end{deluxetable}

\end{document}